\documentclass[aps,preprint]{revtex4-2}

\usepackage{amssymb}
\usepackage{amsmath}
\usepackage{graphicx}
\usepackage{bm}
\usepackage{physics}
\usepackage{hyperref}
\usepackage{float}
\usepackage{subfig}  
\usepackage{subcaption}
\usepackage{dcolumn} 
\usepackage{caption}
\usepackage{natbib}  
\usepackage{xcolor}
\usepackage{geometry}        
\usepackage{float}           
\usepackage{overpic}         

\begin{document}

\title{Moderate Higher-Order Interactions Enhance Stability While Preserving Basin Structure}

\author{Zheng Wang}
 \affiliation{State Key Laboratory of Mechanics and Control for Aerospace Structures, College of Aerospace Engineering, Nanjing University of Aeronautics and Astronautics, Nanjing 210016, China}

\author{Jinjie Zhu}
\email{jinjiezhu@nuaa.edu.cn}
\affiliation{State Key Laboratory of Mechanics and Control for Aerospace Structures, College of Aerospace Engineering, Nanjing University of Aeronautics and Astronautics, Nanjing 210016, China}

\author{Xianbin Liu}
\email{xbliu@nuaa.edu.cn}
\affiliation{State Key Laboratory of Mechanics and Control for Aerospace Structures, College of Aerospace Engineering, Nanjing University of Aeronautics and Astronautics, Nanjing 210016, China}

\begin{abstract}
Synchronization is a ubiquitous phenomenon in complex systems. The Kuramoto model serves as a paradigmatic framework for understanding how coupled oscillators achieve collective rhythm. Conventional approaches focus on pairwise interactions, but real-world systems frequently involve higher-order couplings among multiple elements. Previous studies have shown that higher-order interactions enrich dynamics but generally shrink the attraction basin of synchronized states, making synchronization harder to achieve. Here, we demonstrate this picture is incomplete. Through systematic analysis of twisted states on ring networks, we identify a moderate coupling regime where higher-order interactions enhance stability without altering basin structure. The relative distribution among twisted states remains constant, yet quasipotential barriers deepen as coupling strengths increase. By measuring mean first passage times, we show both pairwise and higher-order couplings contribute synergistically to enhance stability, consistent with large deviation theory. These findings provide new insights into the role of higher-order interactions in synchronization.
\end{abstract}

\maketitle

\section{Introduction}

Synchronization is a fundamental phenomenon in complex systems, observed across diverse natural and engineered contexts, ranging from fireflies flashing in unison and neuronal activity in the brain to power grid stability and laser phase-locking~\cite{pikovsky2001,strogatz2003,Arenas2008,Boccaletti2018}. Understanding the mechanisms underlying synchronization is essential for both explaining natural phenomena and designing robust engineered systems.

Traditionally, interactions among oscillators have been modeled as pairwise connections, where each coupling involves exactly two elements. This framework has proven successful in capturing fundamental synchronization mechanisms through network-based models~\cite{Arenas2008,RODRIGUES2016}. However, many real-world systems exhibit group interactions that transcend pairwise connections. Neuronal dynamics can be shaped by the coordinated activity of multiple presynaptic inputs~\cite{Silver2010}, social contagion processes often depend on group influences rather than dyadic interactions~\cite{Iacopini2019}, and biochemical reaction networks frequently involve simultaneous interactions among multiple molecular species~\cite{Carletti2020}. These higher-order interactions, where the effect on one element depends on the collective state of multiple others, are increasingly recognized as essential features of complex systems~\cite{Battiston2020,Battiston2021}.

The Kuramoto model~\cite{Kuramoto1984,STROGATZ2001} provides a foundational framework for studying synchronization in coupled oscillators. Originally formulated for all-to-all coupled oscillators with pairwise interactions, extensions to complex network topologies have revealed rich phenomena such as cluster synchronization~\cite{RODRIGUES2016}, while non-locally coupled systems exhibit chimera states~\cite{Abrams2004,Panaggio2015}. Recent incorporation of higher-order interactions has profoundly enriched the model's dynamics, with studies showing that strong higher-order coupling generally makes synchronization more difficult to achieve~\cite{Skardal2019,Skardal2020,Millan2020,Leon2024,zhangyz2024}.

A particularly important finding by Zhang et al.~\cite{zhangyz2024} demonstrated that higher-order interactions
 make the attraction basin of the synchronous state smaller but more robust (deeper)  in oscillators on ring hypergraphs.  Similarly, studies on related systems have shown that higher-order interactions reduce the attraction basin while simultaneously enhancing the robustness of synchronized states~\cite{wang2025}. Recent work on hypergraph-coupled oscillators has revealed that weak higher-order interactions can counterintuitively enhance synchronization~\cite{Muolo2025}. These observations reveal a dual effect: higher-order interactions reshape both the geometric structure (basin size) and energetic landscape (basin depth) of synchronization. However, a fundamental question remains: Can higher-order interactions enhance synchronization stability without altering basin structure?

Addressing this question requires distinguishing between two complementary aspects of stability. The basin structure determines which initial conditions lead to synchronization, representing a geometric property of state space. The basin depth, characterized by quasipotential barriers, measures how resistant synchronized states are to perturbations, representing an energetic property. While basin structure has been extensively studied~\cite{menck2013,menck2014,zhang2020,zhangyz2024}, the role of quasipotential barriers in systems with higher-order interactions remains largely unexplored. The quasipotential framework, developed by Freidlin and Wentzell based on large deviation theory~\cite{freidlin1998}, provides a rigorous approach to quantify basin depth through mean first passage times (MFPT) under noise perturbations~\cite{Matkowsky1977,gardiner2009,sliusarenko2010}.

In this work, we investigate the stability of twisted states on ring networks with both pairwise and higher-order interactions. Twisted states are spatially structured synchronized patterns characterized by uniform phase winding along the ring. By systematically exploring the parameter space of pairwise coupling strength $\sigma$ and triadic coupling strength $\sigma_\Delta$, we identify a moderate coupling regime where the basin structure remains nearly unchanged while quasipotential barriers systematically deepen. Our findings reveal that both pairwise and higher-order interactions contribute synergistically to enhance stability through energetic mechanisms, offering insights that may inform the design of robust synchronization in engineered systems such as power grids and neural networks, while deepening our understanding of why higher-order structures are prevalent in nature.

The paper is organized as follows. Section~\ref{sec2} introduces the higher-order Kuramoto model on ring networks and defines twisted states. Section~\ref{sec3} analyzes basin structure in the moderate coupling regime. Section~\ref{sec4} presents quasipotential analysis quantifying basin depth. Section~\ref{sec5} concludes with discussion of implications.

\section{\label{sec2}Model and Theoretical Framework}

Consider a network of $n$ weakly coupled, nearly identical limit-cycle oscillators. Under appropriate conditions, the high-dimensional dynamics can be reduced to a phase description through the phase reduction method~\cite{Kuramoto1984}. The evolution of the phase $\theta_j \in \mathbb{S}^1$ of oscillator $j$ is governed by~\cite{leon2025}:
\begin{equation}
\label{eq:general_model}
\dot{\theta}_j = \omega_j + \varepsilon \sum_{k,l=1}^{N} \mathcal{A}_{jkl} \, \Gamma_{jkl}(\theta_k - \theta_j, \theta_l - \theta_j),
\end{equation}
where $\omega_j$ denotes the natural frequency of oscillator $j$, and $\varepsilon$ is a small parameter quantifying the coupling strength. The adjacency tensor $\mathcal{A}_{jkl}$ encodes the network topology: $\mathcal{A}_{jkl} = 1$ if oscillators $j$, $k$, and $l$ participate in a triadic interaction (with $j \neq k \neq l$), and $\mathcal{A}_{jkl} = 0$ otherwise. The coupling function $\Gamma_{jkl}$ depends solely on phase differences, reflecting the phase-only nature of the reduced dynamics. When restricting the coupling function to its first Fourier harmonics, Eq.~\eqref{eq:general_model} yields an extended Kuramoto model that incorporates both pairwise and higher-order interactions.

To obtain concrete insights while maintaining mathematical tractability, we focus on a ring network with a local coupling radius $r$. In this configuration, each oscillator interacts with neighbors within graph distance $r$ along the circular lattice. The dynamics take the form~\cite{zhangyz2024}:
\begin{equation}
\label{eq:ring_model}
\dot{\theta}_i = \omega + \frac{\sigma}{2r} \sum_{j=i-r}^{i+r} \sin(\theta_j - \theta_i) 
+ \frac{\sigma_\Delta}{2r(2r-1)} \sum_{\substack{j=i-r \\ j \neq i}}^{i+r} \sum_{\substack{k=i-r \\ k \neq i, k \neq j}}^{i+r} \sin(\theta_j + \theta_k - 2\theta_i).
\end{equation}
 Here, $\omega$ represents the common natural frequency of identical oscillators, which we set to zero through a coordinate transformation. The parameter $\sigma$ controls the pairwise coupling strength, normalized by the degree $2r$ to ensure consistent scaling across different coupling radii. Similarly, $\sigma_\Delta$ governs the triadic coupling strength, normalized by $2r(2r-1)$—the number of distinct unordered pairs within the local neighborhood. The triadic coupling term has a natural interpretation: oscillator $i$ experiences a torque proportional to $\sin(\theta_j + \theta_k - 2\theta_i)$, which can be viewed as the phase mismatch between oscillator $i$ and the centroid of oscillators $j$ and $k$. Throughout our numerical investigations, we adopt $n = 83$ oscillators with coupling radius $r = 2$, following the conventions of previous studies~\cite{delabays2017,zhang2021,zhangyz2024}, where $r = 2$ is the minimum radius allowing sufficiently rich triadic interactions.

To characterize the collective dynamics, we introduce a local coherence measure for each oscillator $i$~\cite{zhangyz2024}:
\begin{equation}
\label{eq:local_order}
Z_i = \frac{1}{2r+1} \sum_{j=i-r}^{i+r} e^{i\theta_j},
\end{equation}
where the modulus $|Z_i| \in [0,1]$ quantifies local phase coherence within oscillator $i$'s interaction neighborhood. The value $|Z_i|$ approaches unity when oscillator $i$ and its neighbors exhibit similar phases and vanishes for uniformly distributed phases. We classify the oscillator $i$ as ordered if $|Z_i| \geq \rho_c$, where $\rho_c = 0.85$ serves as an empirically determined threshold that reliably distinguishes coherent regions from incoherent regions. The global order parameter $R$ measures the fraction of ordered oscillators:
\begin{equation}
\label{eq:global_order}
R = \frac{1}{n} \sum_{i=1}^{n} \mathbb{I}(|Z_i| \geq \rho_c),
\end{equation}
where $\mathbb{I}(\cdot)$ denotes the indicator function. This quantity varies between 0 (complete disorder) and 1 (global coherence).

The ring topology admits a family of stationary solutions known as twisted states, characterized by a uniform phase winding along the ring. A $q$-twisted state is defined by:
\begin{equation}
\label{eq:twisted_state}
\theta_i^{(q)} = \frac{2\pi q i}{n} + \Phi, \quad i = 0, 1, \ldots, n-1,
\end{equation}
where $q \in \mathbb{Z}$ is the winding number and $\Phi \in \mathbb{S}^1$ is an arbitrary global phase arising from rotational symmetry. The winding number quantifies the net phase circulation: traversing the ring once accumulates a total phase change of $2\pi q$. The case $q = 0$ corresponds to the synchronized state, where all oscillators share identical phases. States with $|q| > 0$ describe spatially heterogeneous patterns. Due to the system's invariance under simultaneous phase reversal and direction reversal, twisted states with winding numbers $\pm q$ are dynamically equivalent. Furthermore, only winding numbers in the range $-\lfloor n/2 \rfloor \leq q \leq \lfloor n/2 \rfloor$ yield distinct physical configurations~\cite{zhangyz2024}. For twisted states with low winding numbers relative to system size, $|Z_i| \approx 1$ holds for all oscillators, yielding $R = 1$ and confirming global coherence. Intermediate values $0 < R < 1$ signal chimera-like states featuring coexisting coherent and incoherent domains, while completely disordered states yield $R \approx 0$.

\section{\label{sec3}Basin Structure in the Moderate Coupling Regime}

To understand how higher-order interactions influence the basin structure, we systematically investigate the basins of attraction across different coupling parameters. We estimate basin sizes by simulating the dynamics from $1.6 \times 10^5$ random initial conditions uniformly distributed in $[-\pi, \pi)^n$ and recording the fraction of trajectories that converge to each type of attractor. A trajectory is classified as converging to a twisted state if the final configuration satisfies $R = 1$ after a sufficiently long integration time.

\begin{figure}[htbp]
\centering
\captionsetup{justification=raggedright, singlelinecheck=false}
\includegraphics[width=0.55\textwidth]{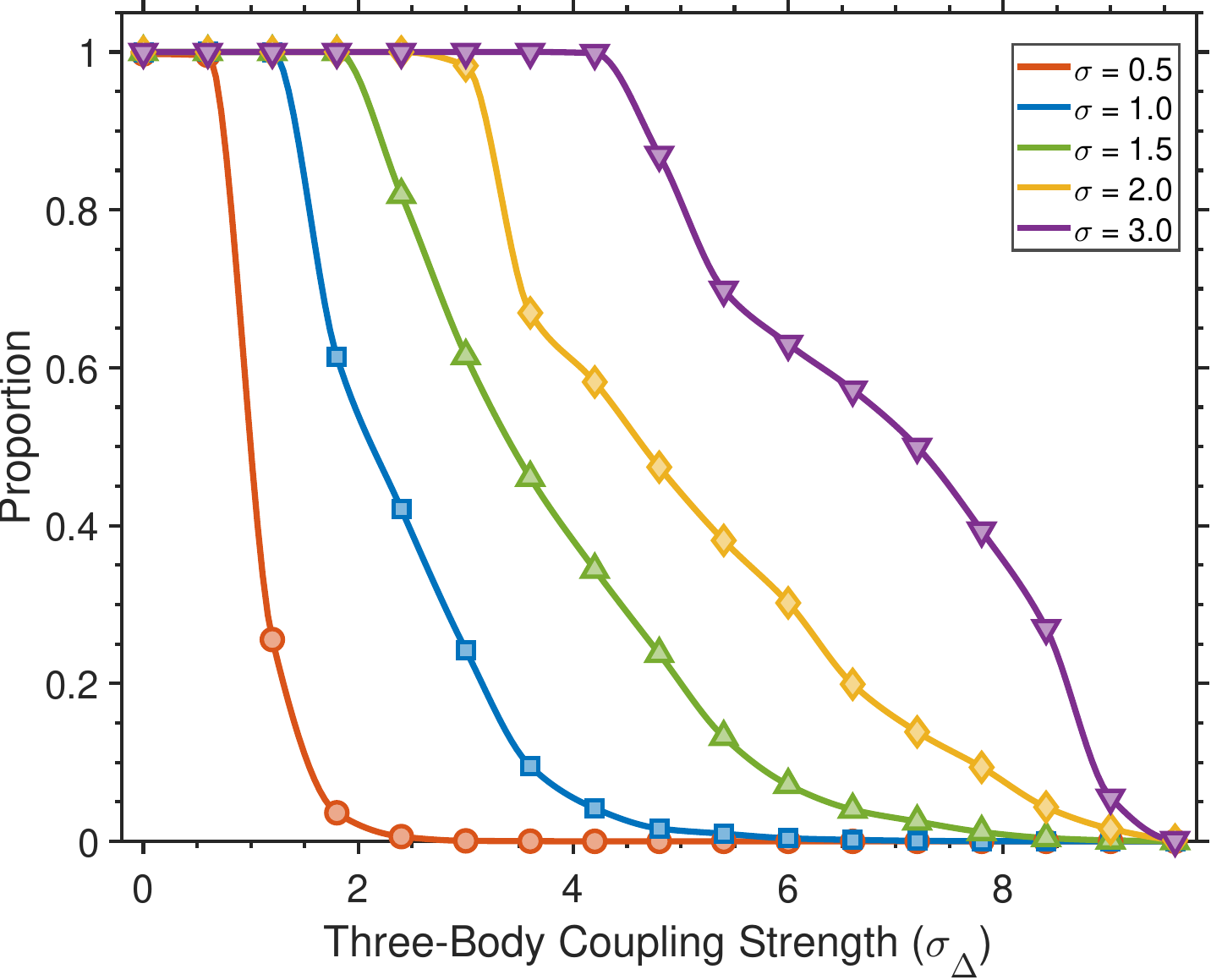}
\caption{Proportion of state space occupied by twisted states ($R = 1$) as a function of triadic coupling strength $\sigma_\Delta$ for different pairwise coupling strengths. Numerical results based on $1.6 \times 10^5$ random initial conditions.}
\label{fig:1}
\end{figure}

FIG.~\ref{fig:1} shows the proportion of twisted states ($R = 1$) as a function of triadic coupling strength $\sigma_\Delta$ for various pairwise coupling strengths $\sigma$. When triadic coupling is moderate, twisted states dominate the entire state space. However, as $\sigma_\Delta$ exceeds a critical threshold, non-twisted states (chimeras and disordered states with $R < 1$) rapidly emerge and eventually dominate, while the basin of twisted states shrinks dramatically. Crucially, this critical threshold increases with stronger pairwise coupling, indicating that sufficient pairwise coupling is necessary to maintain the twisted-state-dominated regime under higher-order interactions.

\begin{figure}[htbp]
\centering
\captionsetup{justification=raggedright, singlelinecheck=false}
\includegraphics[width=1\textwidth]{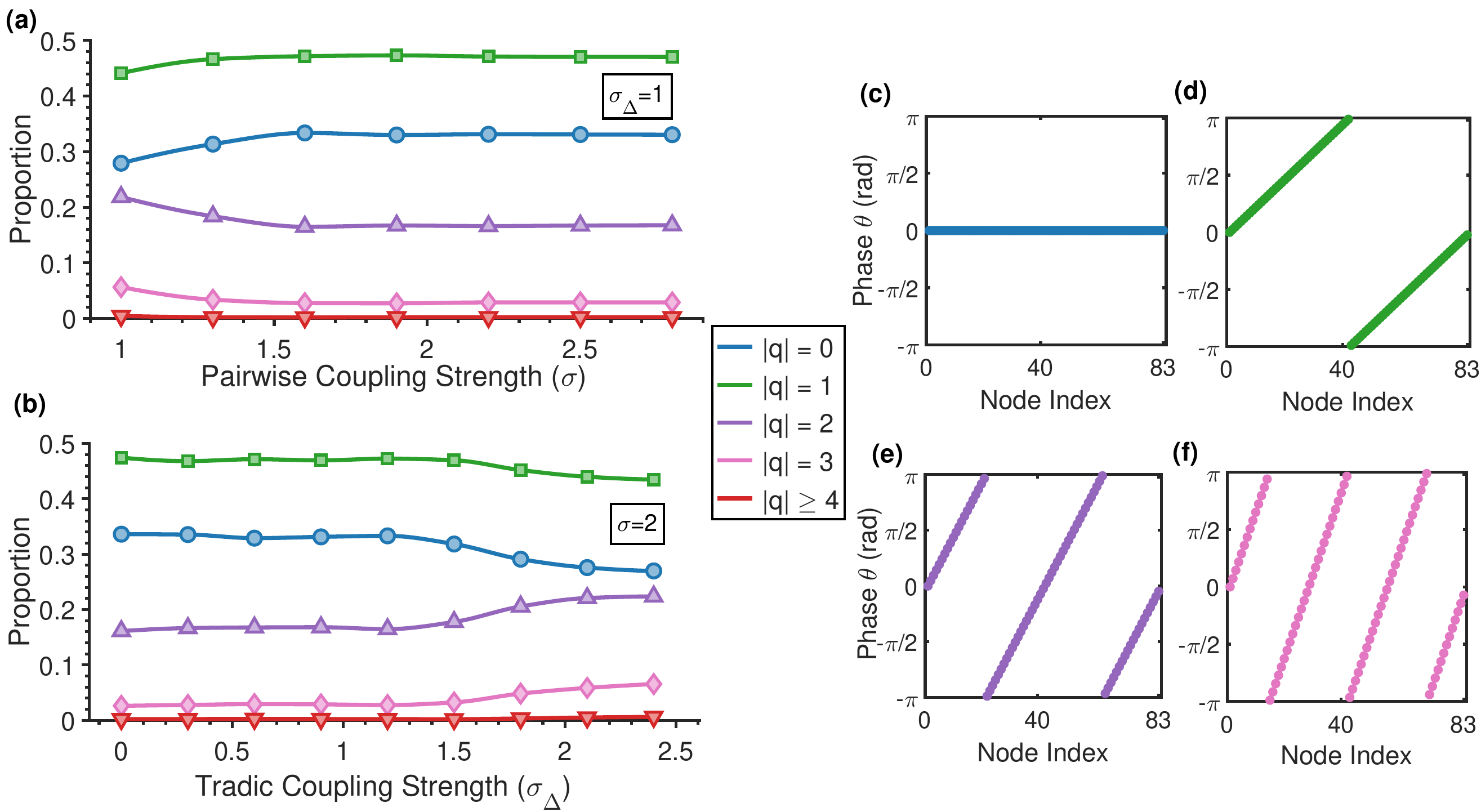}
\caption{Relative proportions of twisted states with different winding numbers in the regime where twisted states dominate the basin structure. (a) Distribution of twisted states as a function of pairwise coupling strength $\sigma $ with fixed $\sigma_\Delta = 1.0$. (b) Distribution as a function of triadic coupling strength $\sigma_\Delta$ with fixed $\sigma = 2.0$. (c-f) Representative phase configurations for twisted states with winding numbers $|q| = 0, 1, 2, 3$, respectively. Numerical results based on $1.6 \times 10^5$ random initial conditions uniformly distributed in $[-\pi, \pi)^n$.}
\label{fig:2}
\end{figure}

Having identified the parameter regime where twisted states dominate, we now examine the internal distribution among different twisted states. FIG.~\ref{fig:2}(a,b) shows the relative proportions of twisted states with different winding numbers in this ordered regime. When either pairwise coupling strength $\sigma$ or triadic coupling strength $\sigma_\Delta$ varies within the twisted state-dominated region, the relative distribution among twisted states remains remarkably stable. States with $|q| = 1$ consistently occupy the largest proportion, followed by the synchronized state ($|q| = 0$) and $|q| = 2$, while higher winding numbers ($|q| \geq 4$) have negligible basin sizes.

These observations reveal a key feature: within the parameter range where non-twisted states are absent, the relative distribution among different twisted states remains nearly constant, largely unaffected by variations in pairwise or triadic coupling strengths. This indicates that, in this parameter region, higher-order interactions have a minimal impact on redistributing basins among twisted states with different winding numbers. FIG.~\ref{fig:2}(c-f) illustrate representative twisted state configurations for different winding numbers.

The combination of FIG.~\ref{fig:1} and \ref{fig:2} defines the moderate higher-order coupling regime that forms the focus of our subsequent analysis: a parameter region where (i) twisted states dominate the basin structure, occupying nearly the entire state space, and (ii) the basin distribution among twisted states remains approximately stationary. In the following, we investigate how higher-order interactions influence the stability of twisted states within this regime.

\section{\label{sec4}Stability Analysis}

To quantify the stability of twisted states beyond deterministic basin analysis, we examine their resilience against noise-induced transitions. We introduce Gaussian white noise into the dynamics by modifying Eq.~\eqref{eq:ring_model}:
\begin{equation}
\label{eq:noise_model}
\dot{\theta}_i = \omega + \frac{\sigma}{2r} \sum_{j=i-r}^{i+r} \sin(\theta_j - \theta_i) 
+ \frac{\sigma_\Delta}{2r(2r-1)} \sum_{\substack{j=i-r \\ j \neq i}}^{i+r} \sum_{\substack{k=i-r \\ k \neq i, k \neq j}}^{i+r} \sin(\theta_j + \theta_k - 2\theta_i) + \sqrt{2}D\xi_i(t),
\end{equation}
where $D$ is the noise intensity and $\xi_i(t)$ are independent Gaussian white noise processes with zero mean and unit variance.

To assess the stability of different states, we employ the mean first passage time (MFPT) as a quantitative measure~\cite{Matkowsky1977,gardiner2009,sliusarenko2010}. The MFPT quantifies the average time required for a stochastic trajectory to escape from one attractor and reach another. For a system starting in state $A$ and transitioning to state $B$, the MFPT is defined as:
\begin{equation}
\label{eq:mfpt}
\tau_e(A \to B) = \mathbb{E}[\inf\{t > 0 : \boldsymbol{\theta}(0) \in A, \boldsymbol{\theta}(t) \in B\}],
\end{equation}
where $\boldsymbol{\theta}$ represents the full phase vector and $\mathbb{E}[\cdot]$ denotes the expectation over noise realizations. The MFPT provides a global measure of stability that complements local stability analysis: longer passage times indicate deeper potential wells and greater resistance to noise-induced escapes.

According to large deviation theory, for sufficiently small noise intensity, the MFPT exhibits exponential scaling with the inverse noise strength~\cite{KRAMERS1940}:
\begin{equation}
\label{eq:kramers}
\tau_e \sim \exp\left(\frac{\Psi}{2D^2}\right),
\end{equation}
where $\Psi$ is the quasipotential barrier height. The quasipotential, introduced by Freidlin and Wentzell for stochastic dynamical systems, generalizes the concept of potential energy to deterministic systems. It is defined through the action functional~\cite{freidlin1998}:
\begin{equation}
\label{eq:quasipotential}
\Psi(A, B) = \inf_{T>0} \inf_{\substack{\phi(0) \in A \\ \phi(T) \in B}} \int_0^T \frac{1}{2}|\dot{\phi}(t) - \mathbf{f}(\phi(t))|^2 dt,
\end{equation}
where $\mathbf{f}(\boldsymbol{\theta})$ represents the deterministic vector field on the right-hand side of Eq.~\eqref{eq:ring_model}, and the infimum is taken over all paths $\phi(t)$ connecting states $A$ and $B$. The quasipotential quantifies the minimum ``cost'' required to deviate from the deterministic trajectory, with rare transitions following optimal paths that minimize this action.

Taking the logarithm of Eq.~\eqref{eq:kramers} yields a linear relationship:
\begin{equation}
\label{eq:linear_scaling}
\ln(\tau_e) \sim \frac{\Psi}{2D^2} = k \cdot \frac{1}{D^2},
\end{equation}
where the slope $k = \Psi/2$ is proportional to the barrier height. By measuring $\tau_e$ for different noise intensities and fitting $\ln(\tau_e)$ versus $1/D^2$, we can extract relative quasipotential barriers and compare the stability of different states.

To measure the MFPT for transitions between twisted states, we implement the following protocol. For transitions from the synchronized state ($|q| = 0$) to non-zero winding number states, we initialize the system in the synchronized configuration and evolve it under noisy dynamics (Eq.~\eqref{eq:noise_model}). At regular intervals, we temporarily remove the noise and allow the system to evolve deterministically for 100 time units (long enough to ensure convergence to a twisted state). We then compute the winding number of the resulting configuration. If the winding number differs from zero, we record a successful escape event and measure the elapsed time as one realization of the first passage time. Similarly, for transitions from non-synchronized twisted states to the synchronized state, we start from random initial conditions that converge to twisted states with $|q| \neq 0$ (i.e., initial conditions within the attraction basin of non-synchronized twisted states), and use the same noise-removal protocol to detect when the system has reached the synchronized state ($|q| = 0$). The MFPT is obtained by averaging over multiple realizations.

\begin{figure}[htbp]
\centering
\captionsetup{justification=raggedright, singlelinecheck=false}
\includegraphics[width=1\textwidth]{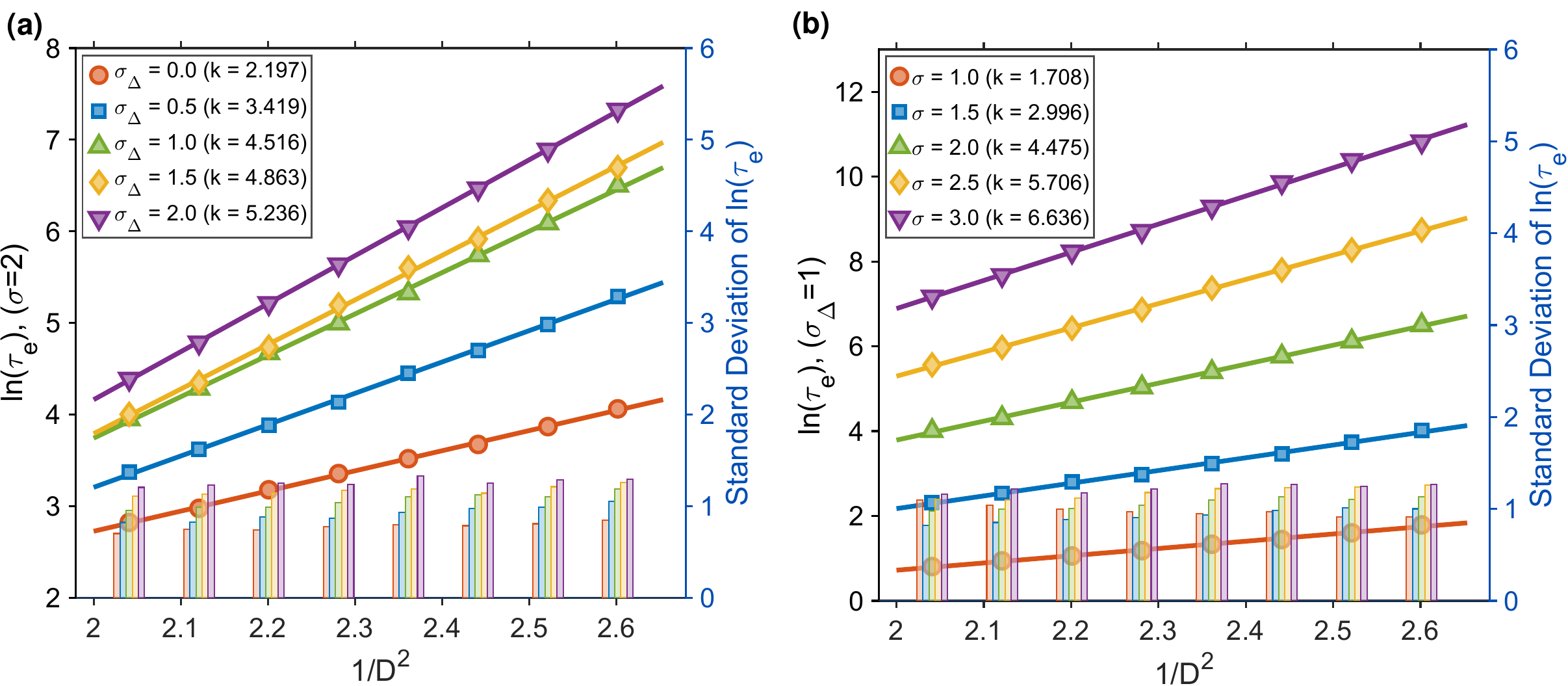}
\caption{Linear fitting of mean first passage time versus noise intensity. (a) Transitions from synchronized state ($|q| = 0$) to non-zero winding states with varying $\sigma_\Delta$ at fixed $\sigma = 2.0$. (b) Transitions from synchronized state to non-synchronized twisted states with varying $\sigma$ at fixed $\sigma_\Delta = 1.0$. Here, $k$ (shown in legend) denotes the slope of the fitted line. The colored bars represent the standard deviations of $\ln(\tau_e)$ for each corresponding data series. Parameters: 1600 sample trajectories.}
\label{fig:3}
\end{figure}

FIG.~\ref{fig:3}(a) shows $\ln(\tau_e)$ as a function of $1/D^2$ for transitions from the synchronized state to non-zero winding number states, with varying triadic coupling strengths $\sigma_\Delta$ at fixed $\sigma = 2.0$. The clear linear relationships confirm the exponential scaling (Eq.~\eqref{eq:kramers}) predicted by large deviation theory. The slope $k$ increases systematically with $\sigma_\Delta$, indicating that higher-order interactions deepen the quasipotential well of the synchronized state, making it increasingly resistant to noise-induced escapes. FIG.~\ref{fig:3}(b) presents analogous results for transitions from the synchronized state to non-synchronized twisted states, with varying pairwise coupling strengths $\sigma$ at fixed $\sigma_\Delta = 1.0$. The slopes reveal that increasing pairwise coupling strength also deepens the quasipotential well, enhancing the stability of the synchronized state against transitions to twisted configurations.

\begin{figure}[htbp]
\centering
\captionsetup{justification=raggedright, singlelinecheck=false}
\includegraphics[width=1\textwidth]{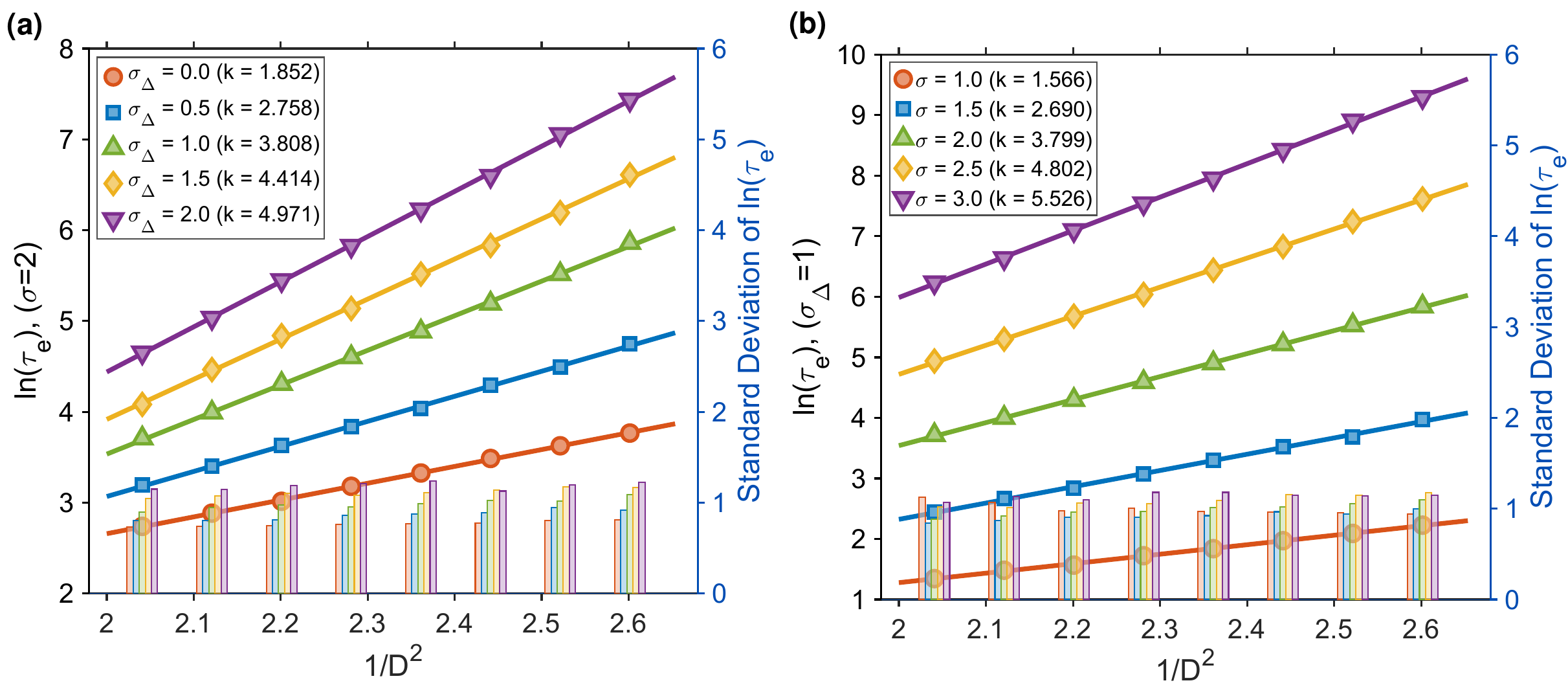}
\caption{Linear fitting of mean first passage time versus noise intensity for transitions from non-synchronized twisted states to synchronized state. (a) Varying triadic coupling strengths $\sigma_\Delta$ at fixed $\sigma = 2.0$. (b) Varying pairwise coupling strengths $\sigma$ at fixed $\sigma_\Delta = 1.0$. Here, $k$ (shown in legend) denotes the slope of the fitted line. The colored bars represent the standard deviations of $\ln(\tau_e)$ for each corresponding data series. Parameters: 1600 sample trajectories.}
\label{fig:4}
\end{figure}

The exponential scaling observed in both panels validates the applicability of quasipotential theory to our high-dimensional, non-gradient system. The linear fits allow us to extract relative well depths, providing a quantitative measure of how coupling parameters reshape the stability landscape. These results demonstrate that, in the moderate coupling regime, both pairwise and higher-order interactions deepen the quasipotential well and enhance the stability of the synchronized state.

\begin{figure}[htbp]
\centering
\captionsetup{justification=raggedright, singlelinecheck=false}
\includegraphics[width=1\textwidth]{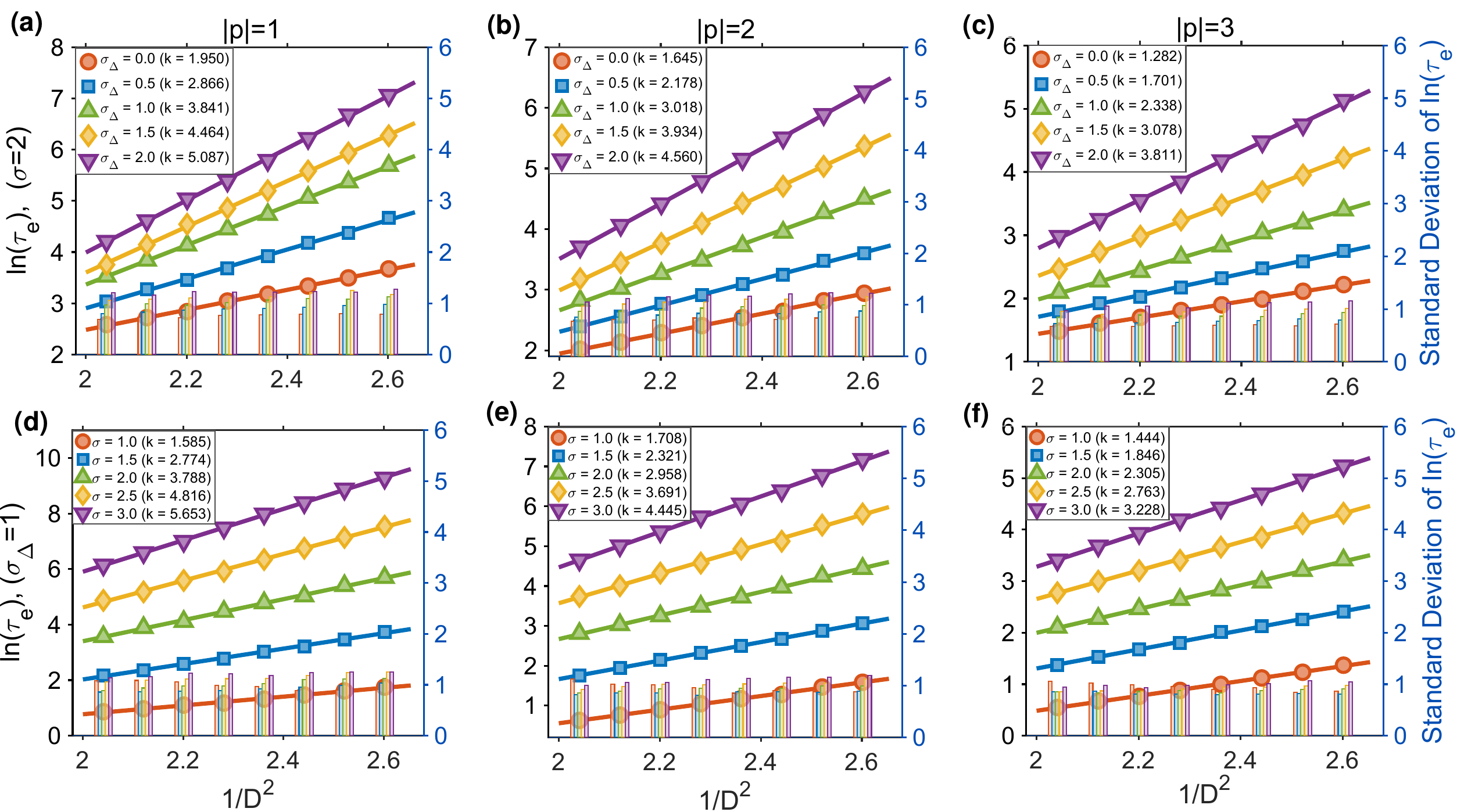}
\caption{Linear fitting of mean first passage time versus noise intensity for transitions from specific twisted states to all other states. Panels (a-c) show results at fixed $\sigma = 2.0$ for twisted states with winding numbers $|p| = 1, 2, 3$, respectively, with varying $\sigma_\Delta$. Panels (d-f) show results at fixed $\sigma_\Delta = 1.0$ for $|p| = 1, 2, 3$, respectively, with varying $\sigma$. Here, $k$ (shown in legend) denotes the slope of the fitted line. The colored bars represent the standard deviations of $\ln(\tau_e)$ for each corresponding data series.  Parameters: 1600 sample trajectories.}
\label{fig5}
\end{figure}

We now examine the reverse transitions from non-zero winding number states to synchronization in FIG.~\ref{fig:4}. In panel (a), with fixed $\sigma = 2.0$, increasing $\sigma_\Delta$ elevates the slopes, indicating that higher-order interactions also stabilize twisted configurations by deepening their quasipotential wells. Similarly, panel (b) shows that stronger pairwise coupling $\sigma$ at fixed $\sigma_\Delta = 1.0$ has the same effect. These complementary results reveal that both coupling mechanisms enhance the stability of both synchronized and twisted states, resulting in a multistable landscape with deep quasipotential wells in the moderate coupling regime.

The results in FIG.~\ref{fig:4} measure transitions from any non-synchronized twisted state to the synchronized state, averaging over different winding numbers. To further elucidate the stability of individual twisted states and their connection to these averaged results, we examine transitions from specific twisted states with winding numbers $|p| = 1, 2, 3$ to all other states. FIG.~\ref{fig5} presents the results: panels (a), (b), and (c) show transitions at fixed $\sigma = 2.0$ for $|p| = 1, 2, 3$, respectively, while panels (d), (e), and (f) show the corresponding results at fixed $\sigma_\Delta = 1.0$. Across all winding numbers, the slopes systematically increase with both pairwise and triadic coupling strengths, confirming that both coupling mechanisms synergistically enhance stability by deepening quasipotential wells. Moreover, comparing across panels reveals a clear trend: twisted states with higher winding numbers exhibit smaller slopes, indicating shallower quasipotential wells and thus lower stability. This observation explains the averaged behavior in FIG.~\ref{fig:4} and aligns with the physical intuition that states with stronger spatial phase gradients are energetically less favorable and more vulnerable to perturbations.

\section{\label{sec5}Conclusion and Discussion}

In this work, we have investigated the stability of twisted states in coupled phase oscillators with moderate higher-order interactions on ring networks. Through systematic basin structure analysis and quasipotential analysis via mean first passage times, we have uncovered key findings regarding how higher-order interactions influence synchronization stability.

Our investigation reveals that, within the moderate higher-order coupling regime, both pairwise and higher-order interactions contribute synergistically to enhance stability. Specifically, in the parameter region where non-twisted states are absent, we find that: (i) the basin structure remains largely preserved, with the relative distribution among different twisted states staying nearly constant as coupling strengths vary; (ii) the quasipotential wells systematically deepen with increasing $\sigma$ or $\sigma_\Delta$, as evidenced by the exponential scaling of $\ln(\tau_e)$ with $1/D^2$ predicted by Freidlin-Wentzell theory. This demonstrates a mechanism whereby higher-order interactions enhance stability by deepening quasipotential wells independently of basin reorganization.

The quasipotential perspective employed in this study extends beyond traditional basin stability analysis. While basin size characterizes which initial conditions lead to synchronization, quasipotential depth measures the stability of the synchronized state. Our results show that in the moderate coupling regime, higher-order interactions primarily enhance stability through the latter mechanism rather than the former, providing insights into the nuanced role of coupling strength in determining system stability under random perturbations.

These findings offer insights for understanding complex systems with higher-order structures. In neural networks, moderate levels of higher-order connectivity could enhance the stability of synchronized states while maintaining robust dynamics. In engineered systems such as power grids, understanding the dual role of basin structure and quasipotential depth could inform strategies for maintaining stable synchronization.

In conclusion, we have demonstrated that moderate higher-order interactions enhance synchronization stability by deepening quasipotential wells while preserving basin structure. This work highlights the importance of considering both geometric (basin) and energetic (quasipotential) aspects of stability in systems with higher-order interactions. As the field continues to develop, such integrated perspectives will be crucial for understanding and designing robust synchronization in complex networks.

\section*{Acknowledgments}

JZ acknowledges the support from the National Natural Science Foundation of China (Grant Nos. 12202195 and 12572037). XL thanks the National Natural Science Foundation of China (Grant No. 12172167) for financial support.

\end{document}